\documentclass[10pt,final,twocolumn]{IEEEtran}
\usepackage{caption}
\usepackage[caption=false]{subfig}
\usepackage{amsmath, amsthm, amssymb}
\usepackage{mathtools}
\usepackage{graphicx}
\usepackage{float}
\usepackage{breqn}
\usepackage{bm}
\usepackage{url}
\usepackage{cite}
\usepackage{placeins}

\begin{document}

\newcommand\numberthis{\addtocounter{equation}{1}\tag{\theequation}}
\newcommand*{\TitleFont}{%
      \usefont{\encodingdefault}{\rmdefault}{b}{n}%
      \fontsize{12}{20}%
      \selectfont}
\numberwithin{equation}{section}

\title{Interface-Flattening Transform for EM Field Modeling in Tilted, Cylindrically-Stratified Geophysical Media}
\author{Kamalesh~Sainath and Fernando L. Teixeira
\thanks{The authors are with the ElectroScience Laboratory (ESL), Department of Electrical and Computer Engineering, The Ohio State University (OSU), Columbus, Ohio, USA 43212 (e-mail: $\{$sainath.1@,teixeira@ece.$\}$osu.edu).}
\thanks{This work was supported by the NASA-NSTRF program and by OSC under Grant PAS-0061.}}
\maketitle
\begin{abstract}
We propose and investigate an ``interface-flattening" transformation, hinging upon Transformation Optics (T.O.) techniques, to facilitate the rigorous analysis of electromagnetic (EM) fields radiated by sources embedded in tilted, cylindrically-layered geophysical media. Our method addresses the major challenge in such problems of appropriately approximating the domain boundaries in the computational model while, in a \emph{full-wave} manner, predicting the effects of tilting in the layers. When incorporated into standard pseudo-analytical algorithms, moreover, the proposed method is quite robust, as it is not limited by absorption, anisotropy, and/or eccentering profile of the cylindrical geophysical formations, nor is it limited by the radiation frequency. These attributes of the proposed method are in contrast to past analysis methods for tilted-layer media that often place limitations on the source and medium characteristics. Through analytical derivations as well as a preliminary numerical investigation, we analyze and discuss the method's strengths and limitations.
\end{abstract}
\begin{IEEEkeywords}
Borehole geophysics, electromagnetic analysis, well-logging, stratified media, transformation optics.
\end{IEEEkeywords}
\section{Introduction}
Many applications demand the numerical evaluation of electromagnetic (EM) fields produced by radiators embedded in complex, inhomogeneous environments \cite{chew}. For example, robust (i.e., with respect to environment and source characteristics) computational modeling of EM sensors operating in complex geological formations is necessary both to better understand the effects of the nearby Earth formation inhomogeneity profile and constitutive properties~\cite{anderson1} on the sensor's response (``forward modeling") and to facilitate resistivity profile inversion (e.g., to assess a formation's hydrocarbon productivity potential)~\cite{anderson1}. The repeated use of the numerical modeler as the forward engine in many inversion methods, combined with the electrically large size of many geophysical problems, requires a fast computational scheme that can rigorously model EM phenomena arising from the subsurface environment's dominant geophysical features.

Bridging the gap between robustness and solution speed, a layered-medium geophysical model is commonly employed. In particular, cylindrically-layered environments characterized by parallel, vertically-oriented (i.e., along $z$) interfaces arise quite frequently~\cite{chew, moon}. This layered-medium approximation is oftentimes motivated (when this domain approximation, near the sensor at least, is approximately valid) because in such media one can employ pseudo-analytical methods based upon cylindrical eigenfunctions expansions. These expansions are particularly attractive since they can rigorously account for eccentered, anisotropic, and/or absorptive geophysical media, as well as a broad range of radiation frequencies, with ease~\cite{moon,teixeira18,teixeira7,chew}. However, it is worthwhile asking whether (and how) one can extend upon the types of media (with respect to, for example, spatial inhomogeneity structure) that can be rigorously modeled using these cylinder wave expansions. In particular, removal of the vertical, parallel-layer modeling constraint, which would facilitate modeling of the presence and effects of \emph{tilted} layered media, offers the distinct opportunity to investigate novel EM wave propagation and scattering phenomena that can aid interpretation of collected EM subsurface sensor data. The presence of relative vertical layer tilt can arise, for example, due to mechanical effects or gravitational pull effects either on the tool itself or on the drilling fluid (mud filtrate) invasion zone in the course of deviated drilling~\cite{teixeira7}. However, rigorous modeling of wave propagation and scattering within these types of media, \emph{without} having to place undesirable constraints on the frequency range, source distribution, and material properties,\footnote{These constraints are imposed, for example, in previous methods analyzing EM waves in tilted, planar-layered media \cite{wait,zhang2}.} is encumbered by a challenge on appropriately approximating the boundaries comprising tilted layers.

To overcome this challenge and enable pseudo-analytical-based modeling of tilted, cylindrically-layered media, we propose and explore a strategy based on Transformation Optics (T.O.) \cite{Pendry23062006,MOP:MOP22784,4154658,kwon1,teixeiraJEWA} to obtain material blueprints of annular slabs to coat each of the cylindrical interfaces in a transformed, standard problem\footnote{For example, c.f. the two intermediate layers in Figs. \ref{geom2c}-\ref{geom2d}.} which contains only vertically-oriented, \emph{parallel} interfaces. This interface flattening is carried out by defining a coordinate mesh/metric deformation followed by incorporating the metric deformation into the ambient material properties of the region originally subject to metric stretching~\cite{teixeiraJEWA,kwon1}.
In particular, through redirecting the wave field's power flow (Poynting vector) the derived T.O. coating slabs cause EM waves impinging upon the vertical, parallel interfaces to reflect off and transmit through the coated interfaces \emph{as if} these interfaces were, in fact, tilted.\footnote{To visualize better the power flow redirection mechanism, see for example Figs. 3a-3b in \cite{pendry5} (illustrated therein, however, for the \emph{planar} analog of our proposed method, which by contrast to \cite{pendry5} addresses the cylindrical case).} Once the equivalent problem is obtained, one can then employ standard analytical tools to calculate the fields radiated in a multi-eccentered, cylindrically-stratified medium~\cite{chew,teixeira18,teixeira7}.
\begin{figure}[H]
\centering
\subfloat[\label{geom1a}]{\includegraphics[width=2in]{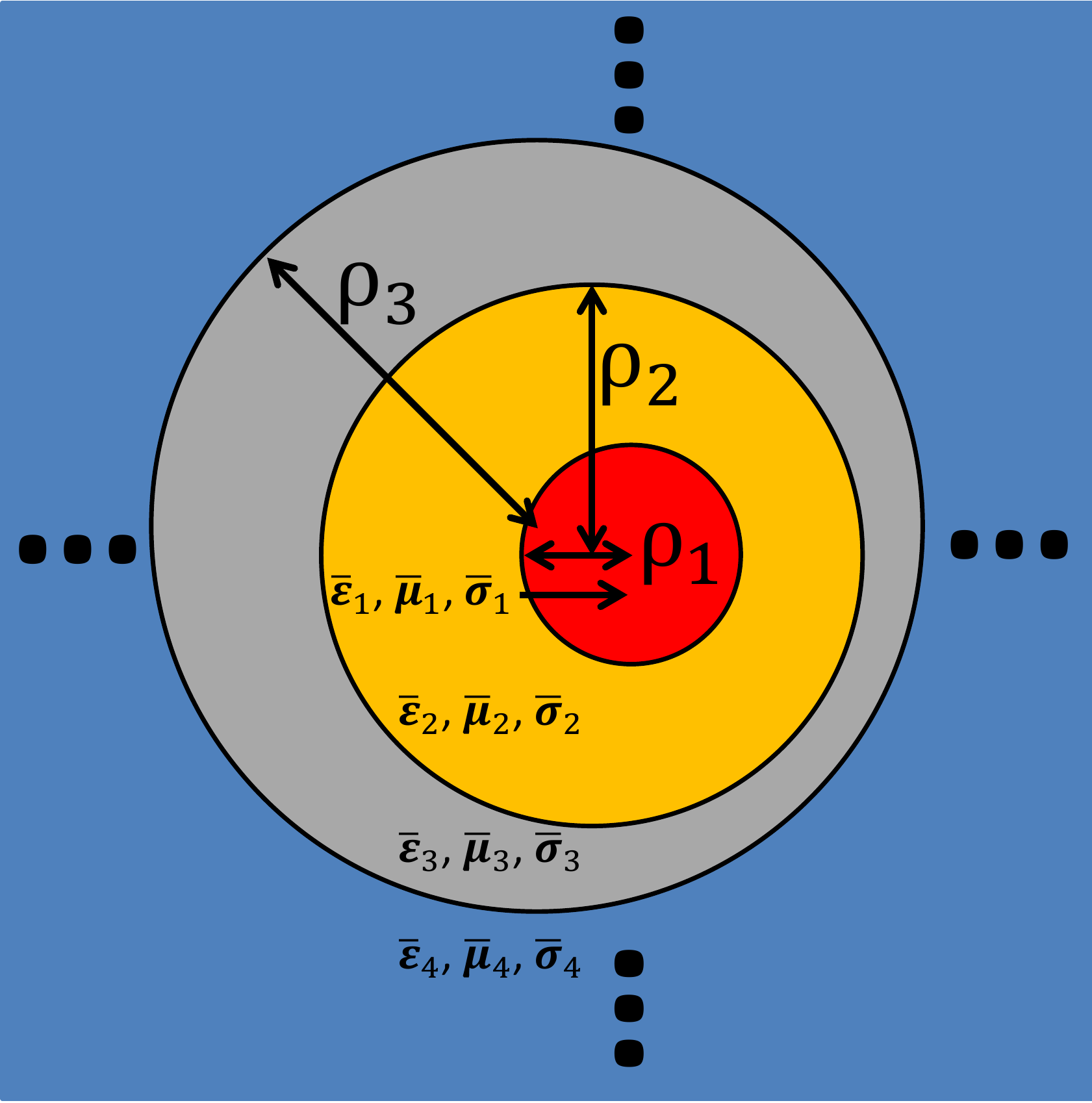}}

\subfloat[\label{geom1b}]{\includegraphics[width=2in]{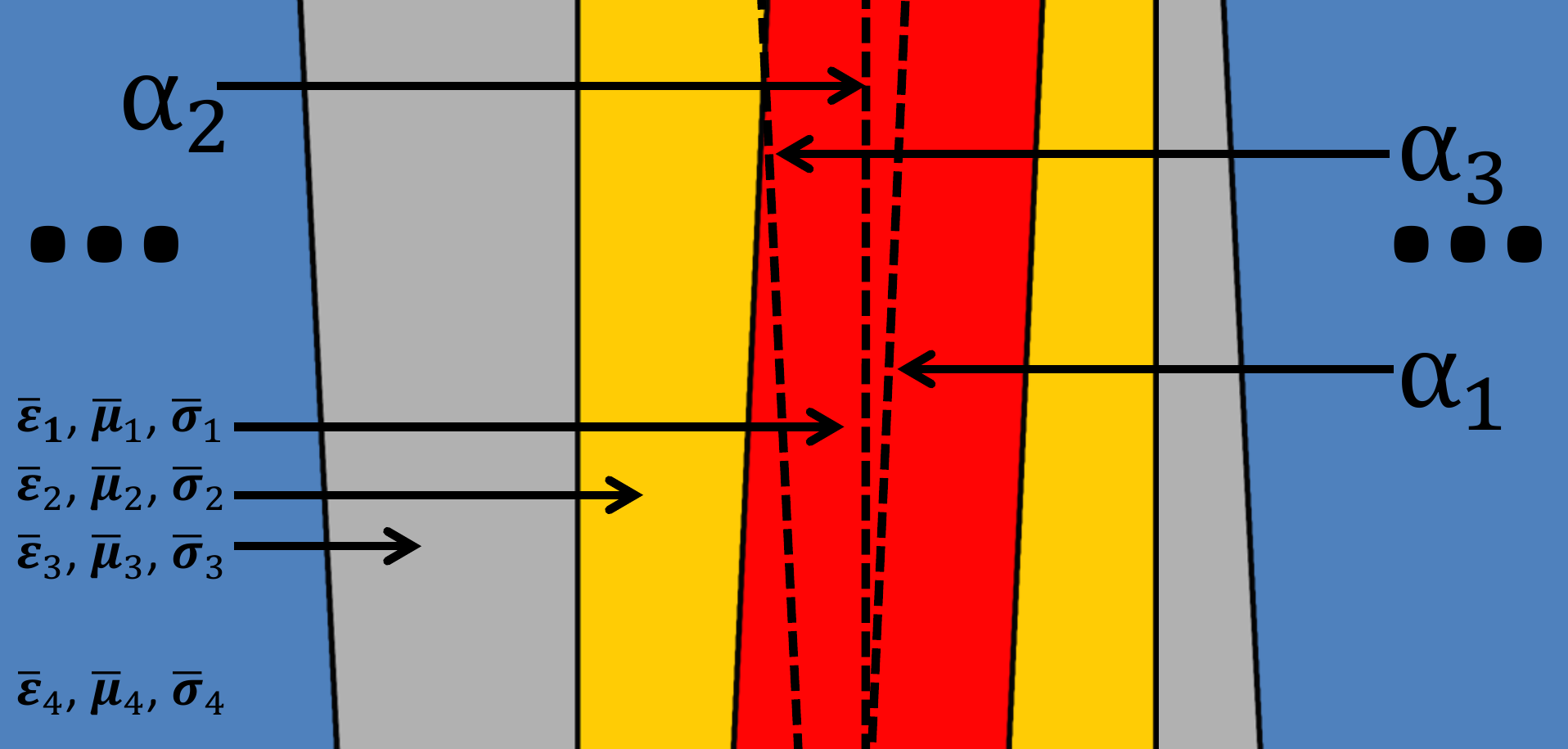}}
\caption{\small Illustration of a tilted, cylindrically-layered formation. Modeled tilting for the $m$th layer ($m\neq N$) is described by effective polar and azimuth tilt orientation angles $\alpha_m$ and $\beta_m$ (resp.) of this $m$th layer's central axis (black dashed lines in Fig. 1b) \cite{anderson1}. For simplicity of illustration, $\beta_1=\beta_2=\beta_3=0^{\circ}$ (i.e., all tilting is parallel to the $xz$ plane).}
\label{geom1}
\end{figure}
\section{Original and Transformed Domains}
\begin{figure}[h]
\centering
\subfloat[\label{geom2a}]{\includegraphics[width=1.5in]{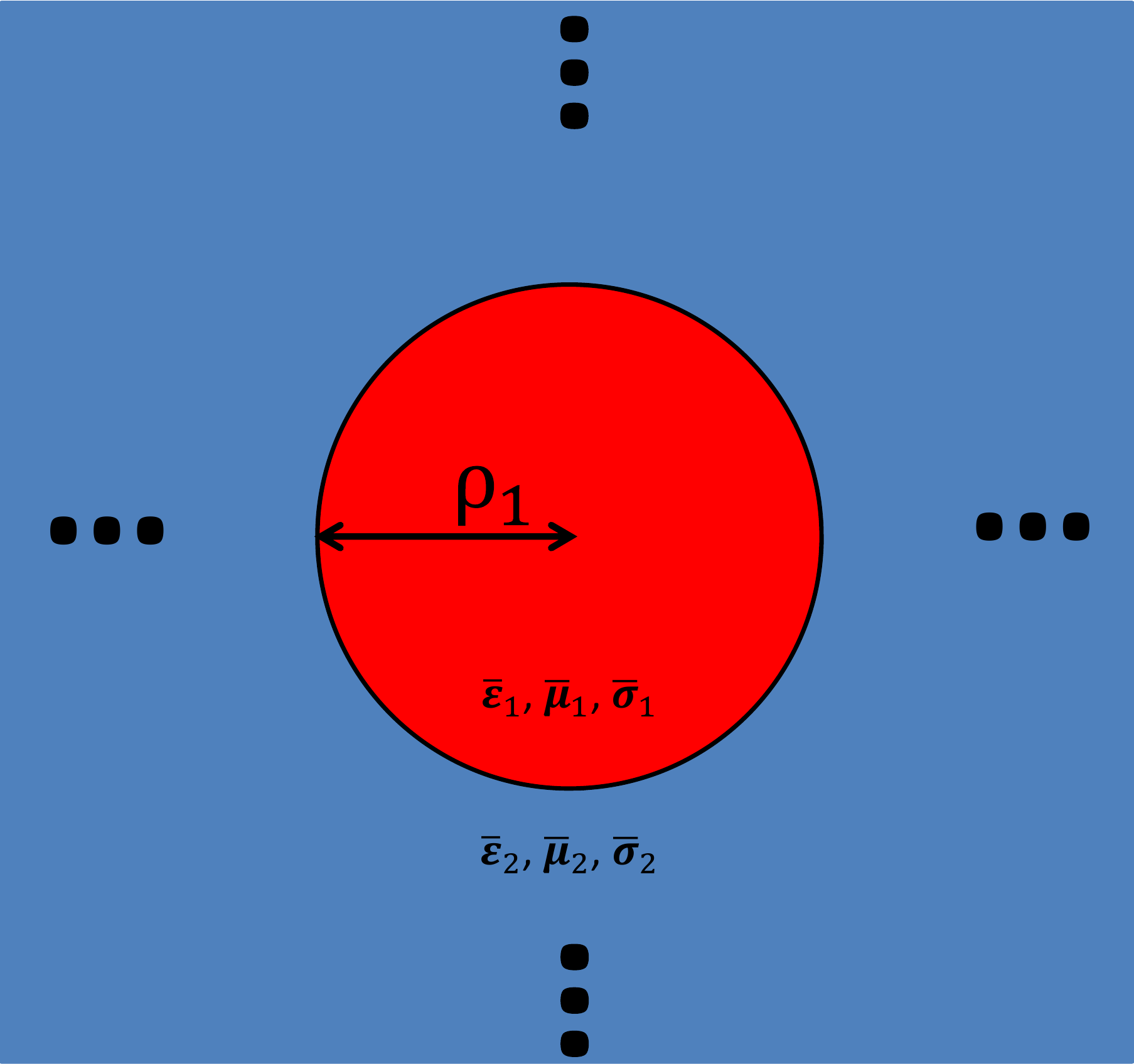}}
\hskip0.2in
\subfloat[\label{geom2b}]{\includegraphics[width=1.72in]{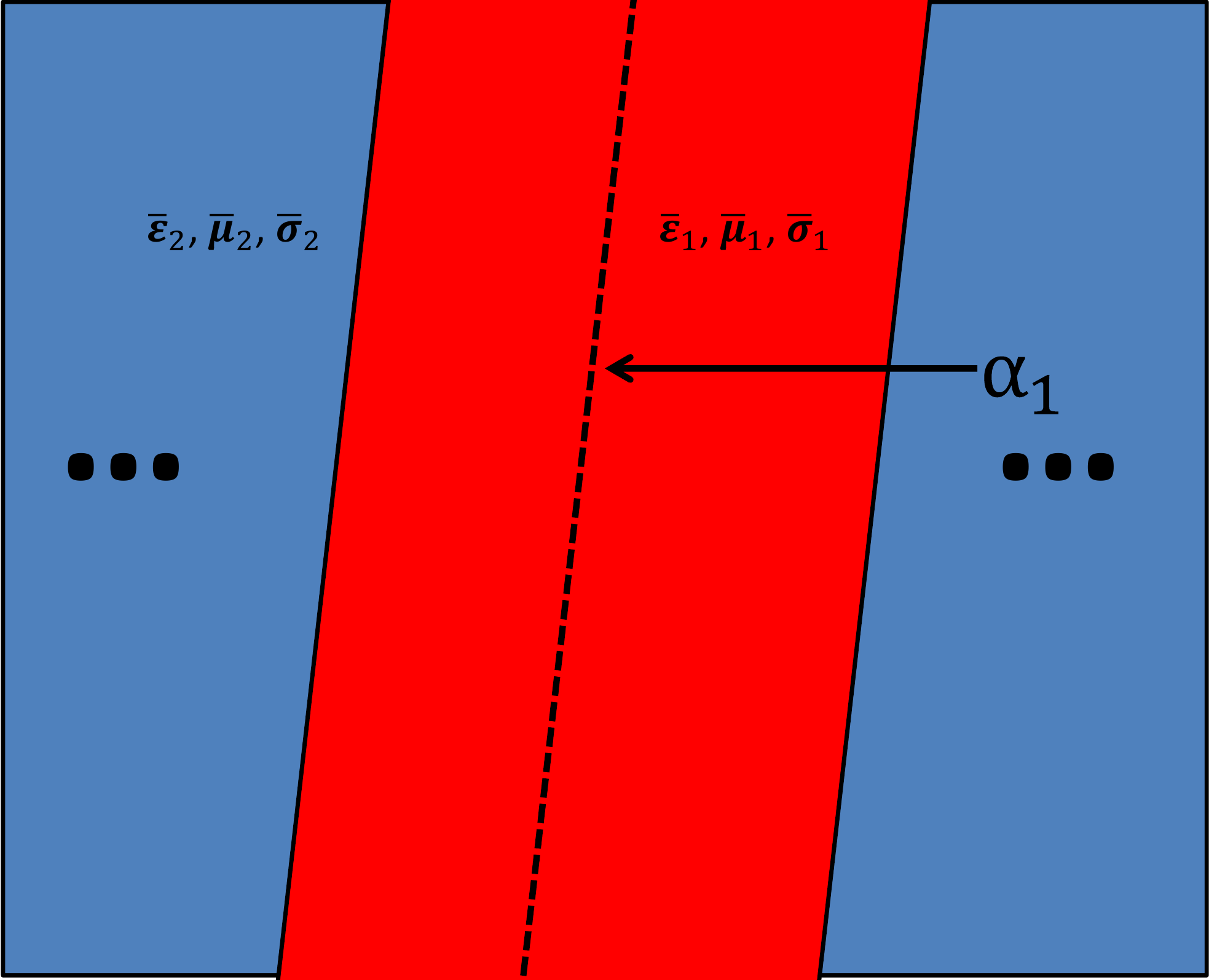}}

\subfloat[\label{geom2c}]{\includegraphics[width=1.5in]{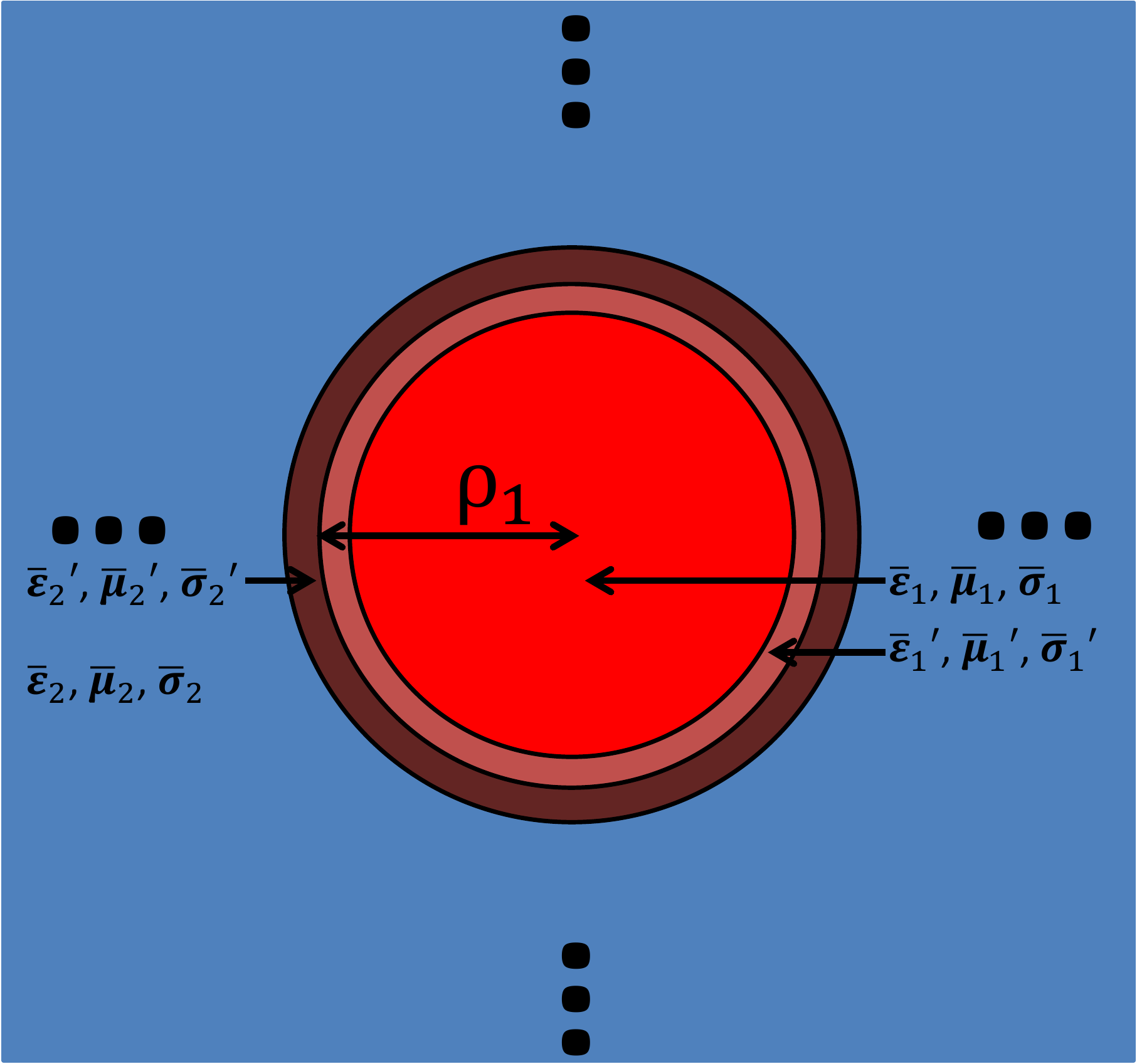}}
\hskip0.2in
\subfloat[\label{geom2d}]{\includegraphics[width=1.72in]{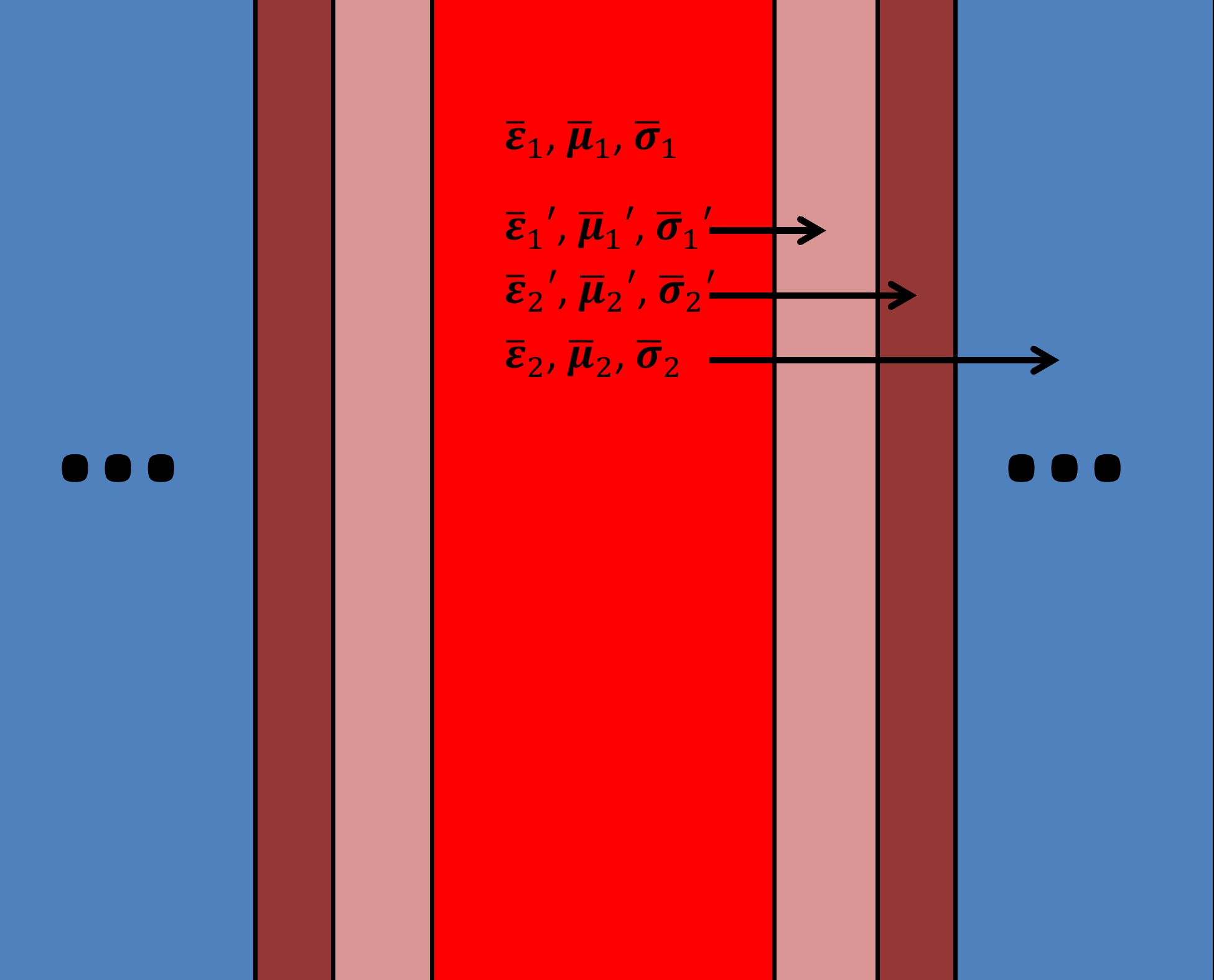}}
\caption{\small Two-layer medium where the T.O.-prescribed media $\boldsymbol{\bar{\tau}}_p'$ ($p$=1,2; $\tau=\{\epsilon,\mu,\sigma \}$), mathematically defined in Section \ref{form},  facilitate domain transformation (from that shown in Figs. \ref{geom2a}-\ref{geom2b}) to a standard domain (shown in Figs. \ref{geom2c}-\ref{geom2d}) of vertically-oriented, cylindrically-stratified media with vertical, parallel interfaces. For simplicity of illustration, $\beta_1=0^{\circ}$.}
\label{geom2}
\end{figure}

Assume $N$ cylindrical layers and $N-1$ interfaces, with layers 1 and $N$ (resp.) referring to the inner-most and outer-most layers.\footnote{\label{glo}To allow for eccentricity in the formations (i.e., the cylindrical layers not sharing the same transverse center), note that we address all aspects of the theoretical derivation of our method with respect to a particular formation's ``local" coordinates. Indeed, we use ``global" coordinates ($x,y,z$) to parameterize the relative positioning of the cylindrical layers, while formation $m$ ($m$=1,2,...,$N$) has ``local" coordinates ($x_m,y_m,z_m=z$) with origin $x_m=y_m=0$ at the $m$th cylinder's center ($x=x_{m,c},y=y_{m,c},z=0$).} Furthermore, the $m$th cylindrical formation (for $m\neq N$) has its exterior interface located at $\rho_m'=\sqrt{(x_m')^2+(y_m')^2}$ meters from its local origin (c.f. footnote \ref{glo}). See Figs. \ref{geom1a} and \ref{geom1b} for an illustration of a typical four-layered geometry at the (resp.) $xy$ plane view (illustrating layer eccentricity) and $xz$ plane view (illustrating relative tilting between the three cylindrical interfaces). To model problems with \emph{relative} interface tilting but employing only vertically-oriented, parallel cylindrical formations, we propose here the use of annular ``coating slabs" on the immediate exterior and immediate interior regions surrounding each vertically-oriented interface. Taking the simple example of a two-layer medium, this can be seen as transforming the original domain (c.f. Figs. \ref{geom2a}-\ref{geom2b}) to one of vertical, parallel interfaces (c.f. Figs. \ref{geom2c}-\ref{geom2d}).
\section{\label{form}Formulation Overview}
We now outline the mathematical derivation of the T.O.-inspired material tensors for the coating slab. First let us examine the $m$th vertical cylinder in its local coordinate system $(x_m,y_m,z_m)$ with interface located at $\rho_m'$, and assume we wish waves to scatter off this interface \emph{as if} its central axis were oriented within the $x_m z_m$ plane by polar angle $\alpha_m$ and azimuthal angle $\beta_m=0^{\circ}$ (i.e., as if the central axis were tilted to point in the direction $\bold{\hat{x}}\sin{\alpha_m}+\bold{\hat{z}}\cos{\alpha_m}$). Now define the following coordinate transformation, applied to the annular slab region $\{(\rho_m'-d) \leq \rho_m \leq (\rho_m'+d) \}$,
\begin{equation}
\tilde{x}_m=x_m, \ \tilde{y}_m=y_m, \ \tilde{z}_m=z_m+(x_m+[x_m'-d])\tan{\alpha_m} \label{transform}
\end{equation}
that redirects the incident field's power flow such that it illuminates the vertically-oriented cylindrical interface \emph{as if} it had the desired tilting described above. This coordinate transformation \eqref{transform}, prescribed only in the $2d$ meter thick annular vicinity of interface $m$, induces a local (effective) distortion of the spatial metric tensor $\bold{\bar{g}}_m$ and associated Jacobian matrix $\bold{\bar{\Lambda}}_m$.\footnote{That is, distorted with respect to undistorted/flat space containing vertical, parallel interfaces.} In particular, $\bold{\bar{\Lambda}}_m$ writes as follows \cite{pendry5}:
\begin{align} \label{lambda}
\boldsymbol{\bar{\Lambda}}_m=\begin{bmatrix} \frac{\partial \tilde{x}_m}{\partial x_m} & \frac{\partial \tilde{y}_m}{\partial x_m} & \frac{\partial \tilde{z}_m}{\partial x_m} \\ \frac{\partial \tilde{x}_m}{\partial y_m}& \frac{\partial \tilde{y}_m}{\partial y_m}& \frac{\partial \tilde{z}_m}{\partial y_m}\\ \frac{\partial \tilde{x}_m}{\partial z_m}& \frac{\partial \tilde{y}_m}{\partial z_m}&\frac{\partial \tilde{z}_m}{\partial z_m} \end{bmatrix}=\begin{bmatrix} 1 & 0 & \tan{\alpha_m} \\ 0& 1& 0\\ 0 & 0&1 \end{bmatrix}
\end{align}
with the inverse metric tensor writing as $\bold{\bar{g}}_m^{-1}=\boldsymbol{\bar{\Lambda}}_m^{\mathrm{T}}\cdot \boldsymbol{\bar{\Lambda}}_m$.\footnote{For reference, more general tilting can be effected through setting, in \eqref{lambda}, $\frac{\partial \tilde{z}_m}{\partial x_m}=\tan{\alpha_m}\cos{\beta_m}$ and $\frac{\partial \tilde{z}_m}{\partial y_m}=\tan{\alpha_m}\sin{\beta_m}$.} To effect computation of Maxwellian fields, as usual one then incorporates the effects of the coordinate transform (embedded within $\bold{\bar{g}}_m$) into the material tensors of the annular region, surrounding the $m$th interface, subject to the coordinate transform.

As an example of incorporating the induced metric tensor into the material properties, for simplicity of illustration assume a two-layer cylindrically-stratified medium (see Figs. \ref{geom2a}-\ref{geom2d}). Letting the inner and outer layers (resp.) initially be uniformly characterized (c.f. Figs. \ref{geom2a}-\ref{geom2b}) by the dielectric, permeability, and conductivity material tensors $\{\boldsymbol{\bar{\epsilon}}_1,\boldsymbol{\bar{\mu}}_1,\boldsymbol{\bar{\sigma}}_1\}$ and $\{\boldsymbol{\bar{\epsilon}}_2,\boldsymbol{\bar{\mu}}_2,\boldsymbol{\bar{\sigma}}_2\}$, and letting $\boldsymbol{\bar{\tau}}_p'=\boldsymbol{\bar{\Lambda}}_1^{\mathrm{T}}\cdot \boldsymbol{\bar{\tau}}_p \cdot \boldsymbol{\bar{\Lambda}}_1$ ($p$=1,2; $\tau=\{\epsilon,\mu,\sigma \}$) represent the corresponding material tensors of the interface-coating annular slabs after incorporation of $\bold{\bar{g}}_1$, one has the following inhomogeneity profile:
\begin{align}
\{\boldsymbol{\bar{\epsilon}}_1,\boldsymbol{\bar{\mu}}_1,\boldsymbol{\bar{\sigma}}_1\}&, \ 0\leq \rho_1 < (\rho_1'-d) \\
\{\boldsymbol{\bar{\epsilon}}_1',\boldsymbol{\bar{\mu}}_1',\boldsymbol{\bar{\sigma}}_1'\}&, \ (\rho_1'-d)\leq \rho_1 < \rho_1' \\
\{\boldsymbol{\bar{\epsilon}}_2',\boldsymbol{\bar{\mu}}_2',\boldsymbol{\bar{\sigma}}_2'\}&, \ \rho_1' \leq \rho_1 < (\rho_1'+d) \\
\{\boldsymbol{\bar{\epsilon}}_2,\boldsymbol{\bar{\mu}}_2,\boldsymbol{\bar{\sigma}}_2\}&, \ (\rho_1'+d) \leq \rho_1 < \infty
\end{align}
The extension to multiple layers follows naturally, as does application of this method to eccentered layers, by performing the transformation around each vertically-oriented interface. Furthermore, recalling the matrix transforming a vector's representation from the cartesian to cylindrical coordinate system \cite{chew}
\begin{equation}
\bold{\bar{U}}_m=\begin{bmatrix} \cos{\phi_m}&\sin{\phi_m} & 0\\ -\sin{\phi_m}&\cos{\phi_m} & 0\\0 &0 & 1\end{bmatrix}
\end{equation}
one has the corresponding cylindrical coordinate representation of the T.O. material tensors $\boldsymbol{\bar{\gamma}}_{p}'=\bold{\bar{U}}_m\cdot \boldsymbol{\bar{\tau}}'_{p}\cdot \bold{\bar{U}}_m^{-1}$ ($\gamma,\tau=\epsilon,\mu,\sigma; m=1,2,...,N-1$; $p=m,m+1$) defined with respect to their cartesian representation counterparts $\{\boldsymbol{\bar{\tau}}'_{p}\}$.
Note that each cylindrical layer, in the new (approximately) equivalent problem consisting of vertically-oriented cylindrical formations with parallel interfaces, is homogeneous (with respect to cartesian coordinates) due to the homogeneous nature of the effective metric distortion.\footnote{That is, $\boldsymbol{\bar{\Lambda}}_m$ within the $m$th interface's annular coating slab region is independent of the cartesian spatial coordinates.} This enables simulations using a standard cylindrical-layer algorithm admitting generally anisotropic, eccentered layers ~\cite{teixeira18} without having to discretize and/or otherwise approximate the coating slab's inhomogeneous material profile.

A drawback in the discussed coordinate transform is that it does \emph{not} result in the prescription of exactly reflectionless T.O. media. From a physical standpoint, to ensure a reflectionless T.O. medium one must not perturb tangential field continuity at the interface adjoining the T.O. medium to the external/ambient medium. Mathematically speaking, one can indeed deform all three coordinates (i.e., both those tangential and normal to the interface) but (in the cartesian case, for example) a necessary condition for reflectionless behavior is that the deformed coordinate $\tilde{x}_m$ can depend only on its un-stretched counterpart ($x$) and the coordinate \emph{normal} to the interface (and likewise for $\tilde{y}_m$ and $\tilde{z}_m$). On the other hand, to ``flatten" the cylindrical interfaces we prescribed a coordinate transform $\tilde{z}_m$ that depends (for example) on $x_m$ in the case of modeling layer tilting in the $x_mz_m$ plane (c.f. \eqref{transform}); however, $x_m$ is \emph{not} uniformly normal to the flattened, vertical interface. It is only so at azimuth angles $\phi_m=0^{\circ}$ and $\phi_m=180^{\circ}$; by contrast, $x_m$ is orthogonal to the interface at $\phi_m=\pm 90^{\circ}$. This results in an azimuth-dependent impedance mismatch. The numerical results presented in the next section investigate this point further.
\section{\label{num}Numerical Results}
\begin{figure}[h]
\centering
\subfloat[\label{geom3a}]{\includegraphics[height=1.7in,width=1.8in]{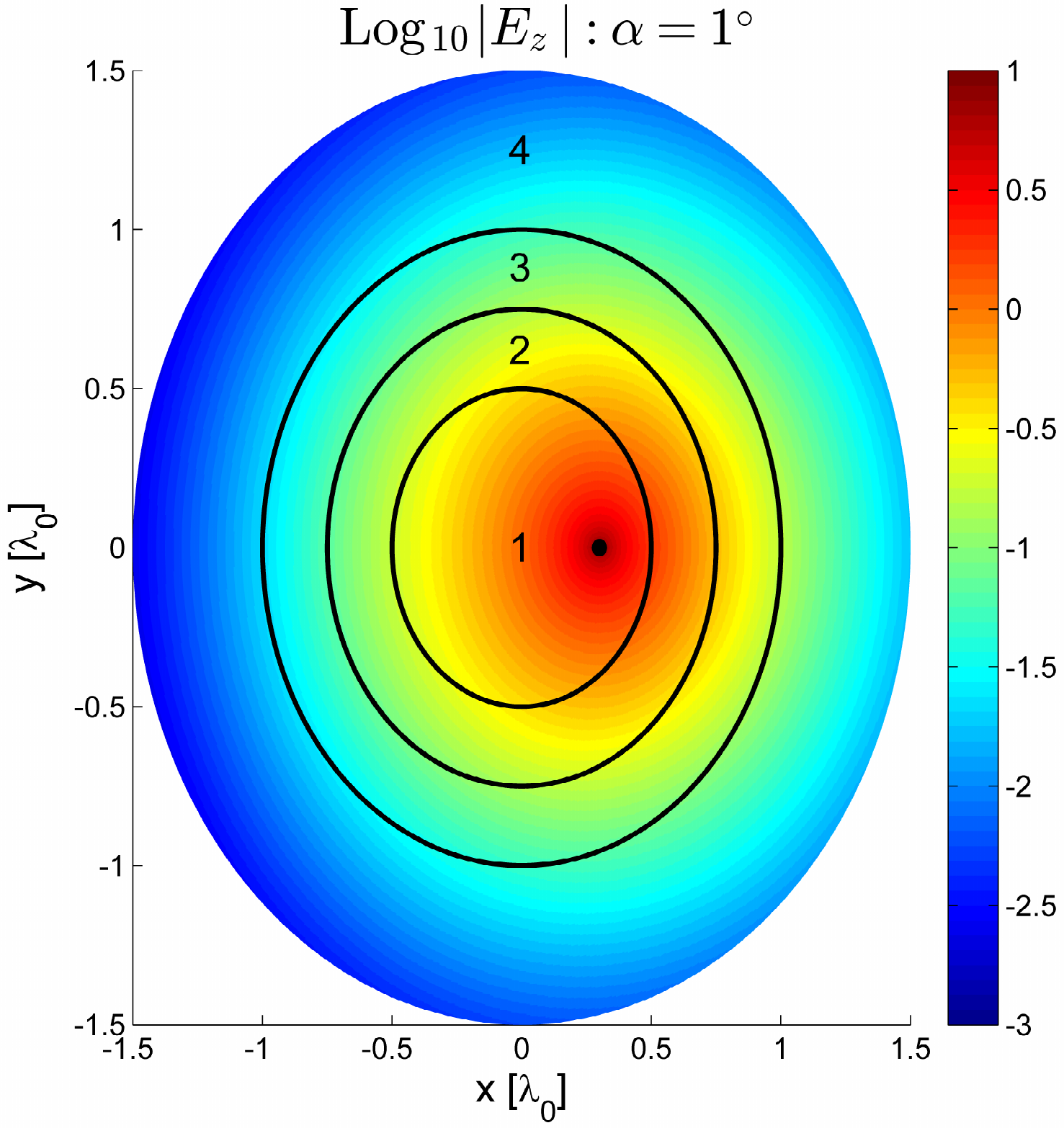}}
\subfloat[\label{geom3b}]{\includegraphics[height=1.7in,width=1.8in]{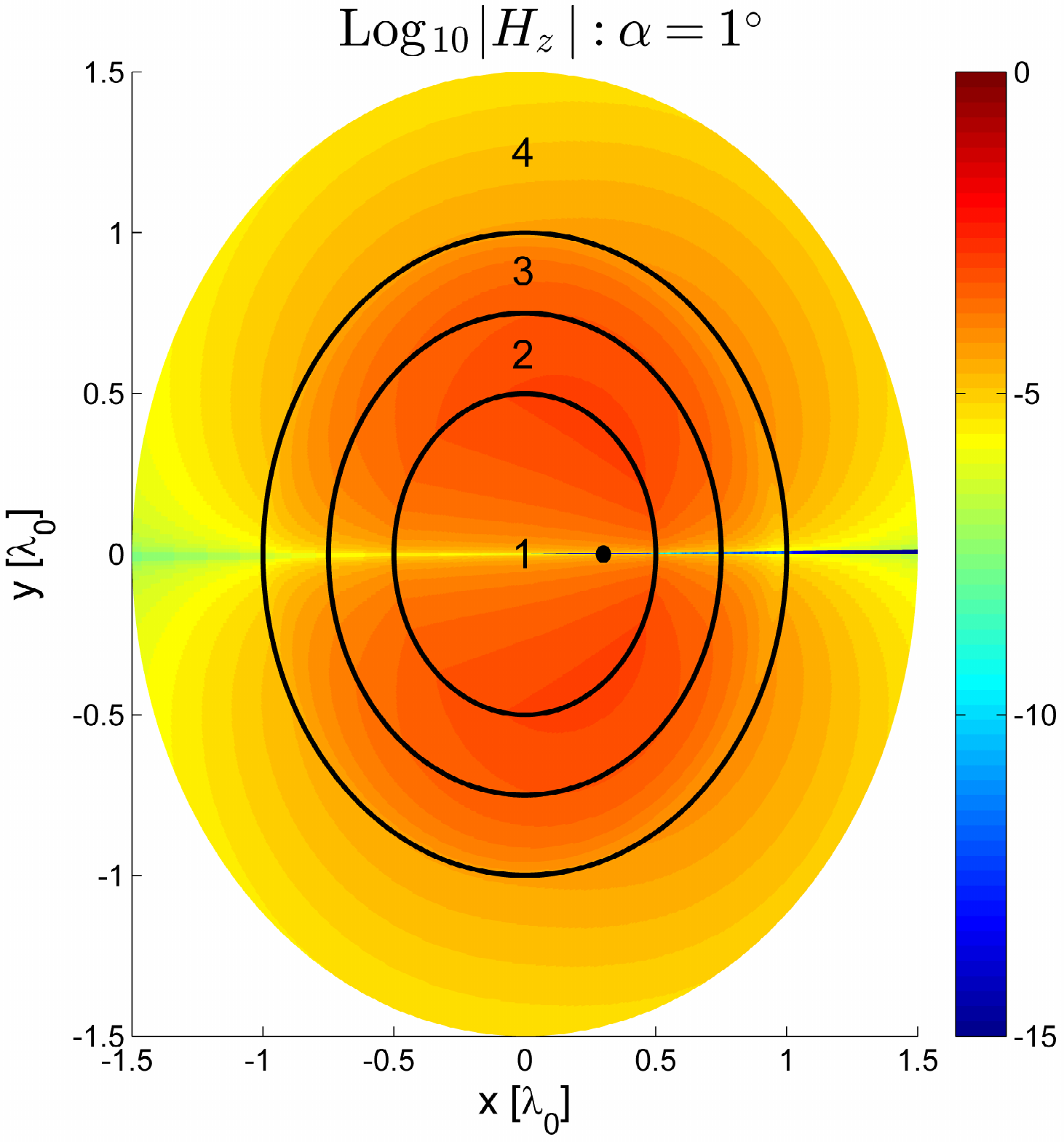}}

\subfloat[\label{geom3c}]{\includegraphics[height=1.7in,width=1.8in]{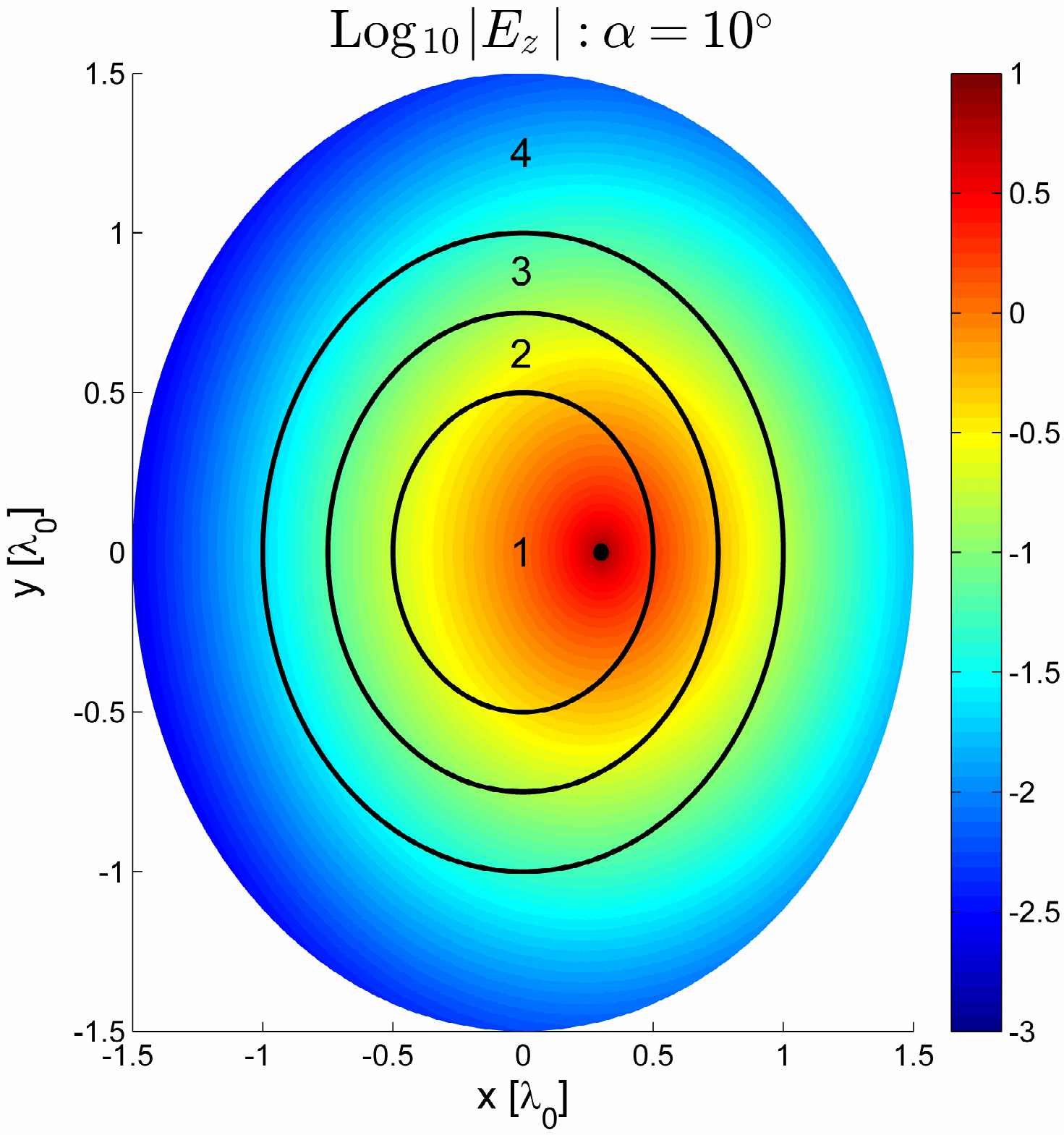}}
\subfloat[\label{geom3d}]{\includegraphics[height=1.7in,width=1.8in]{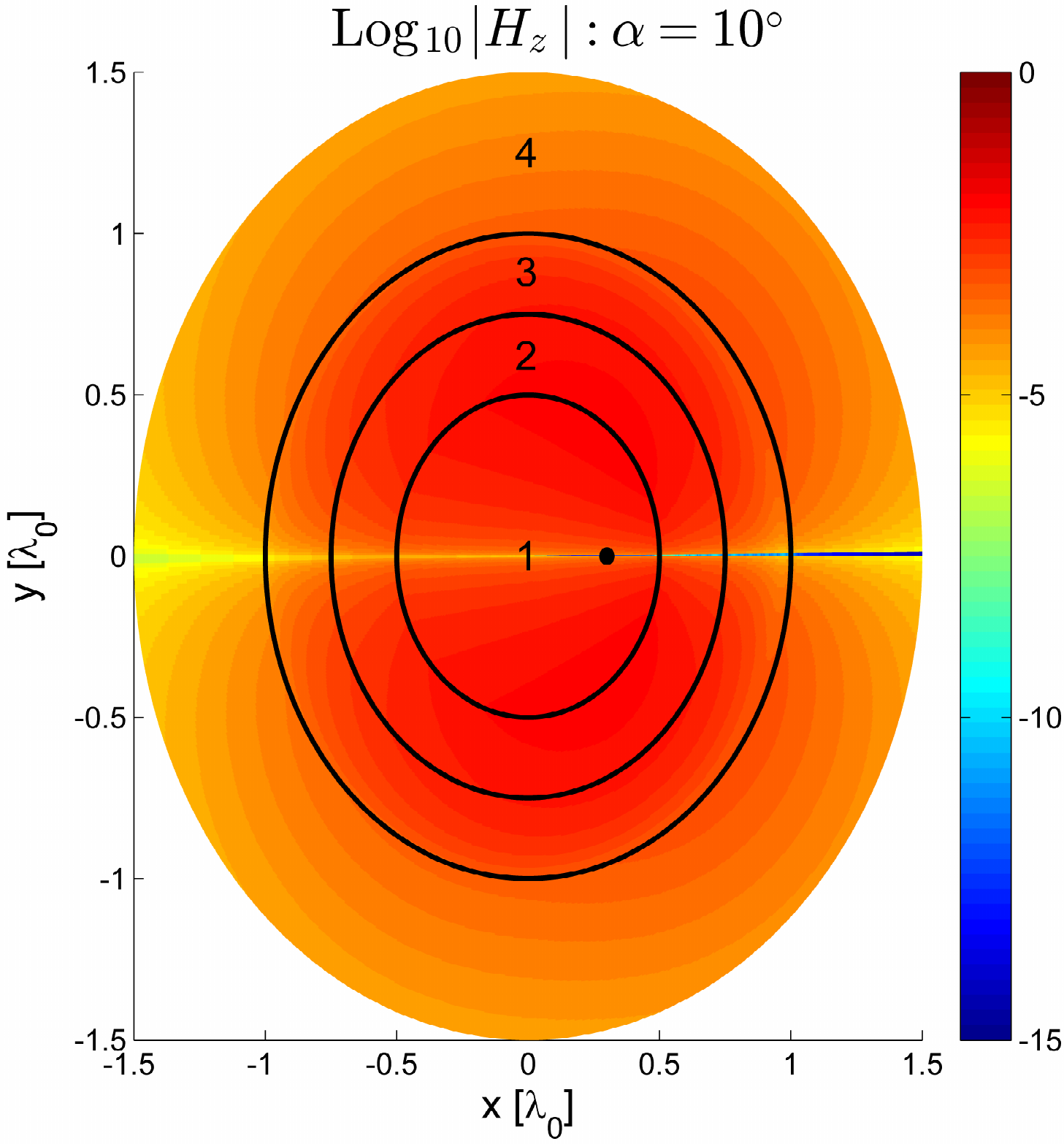}}

\subfloat[\label{geom3e}]{\includegraphics[height=1.7in,width=1.8in]{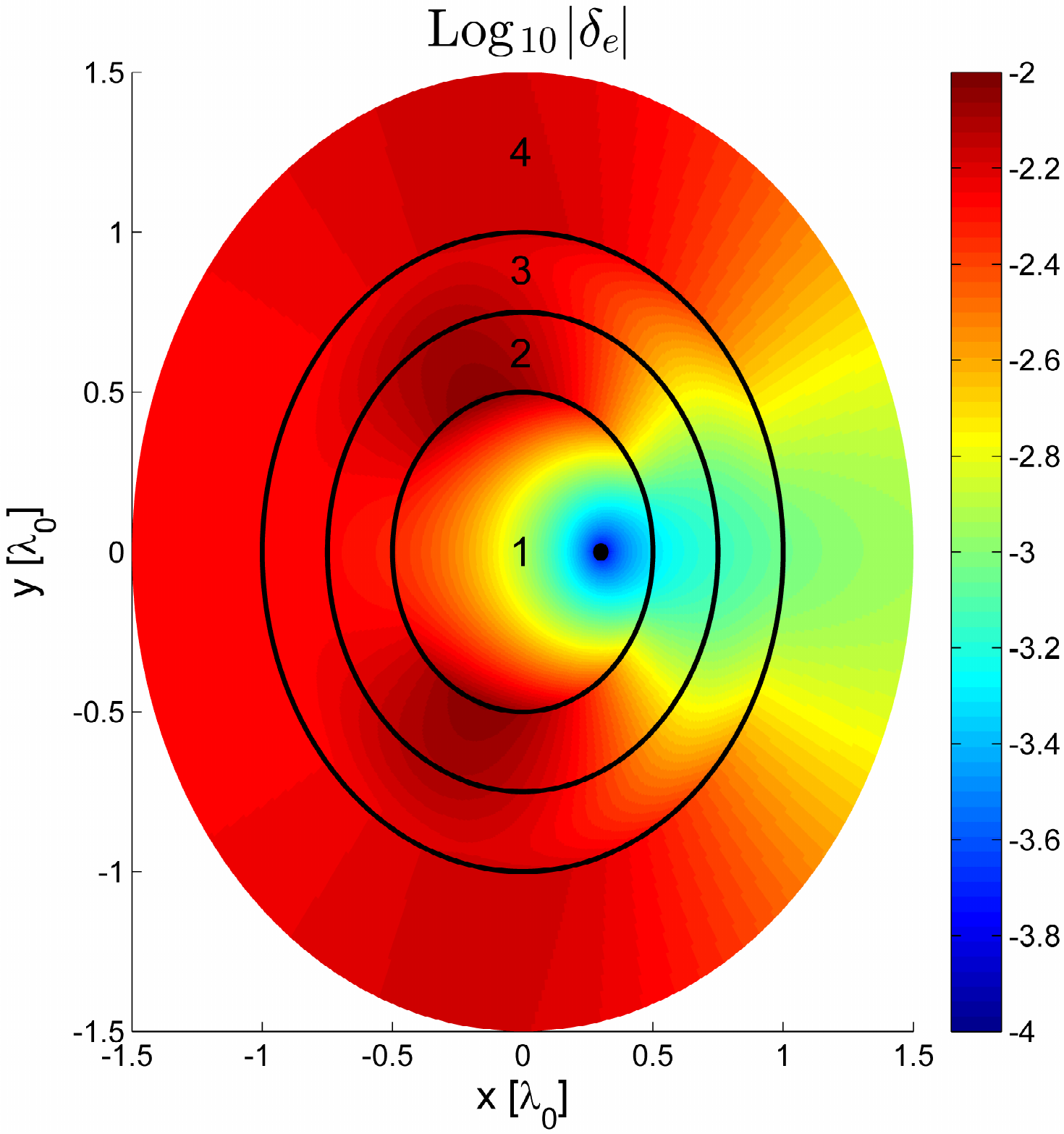}}
\subfloat[\label{geom3f}]{\includegraphics[height=1.7in,width=1.8in]{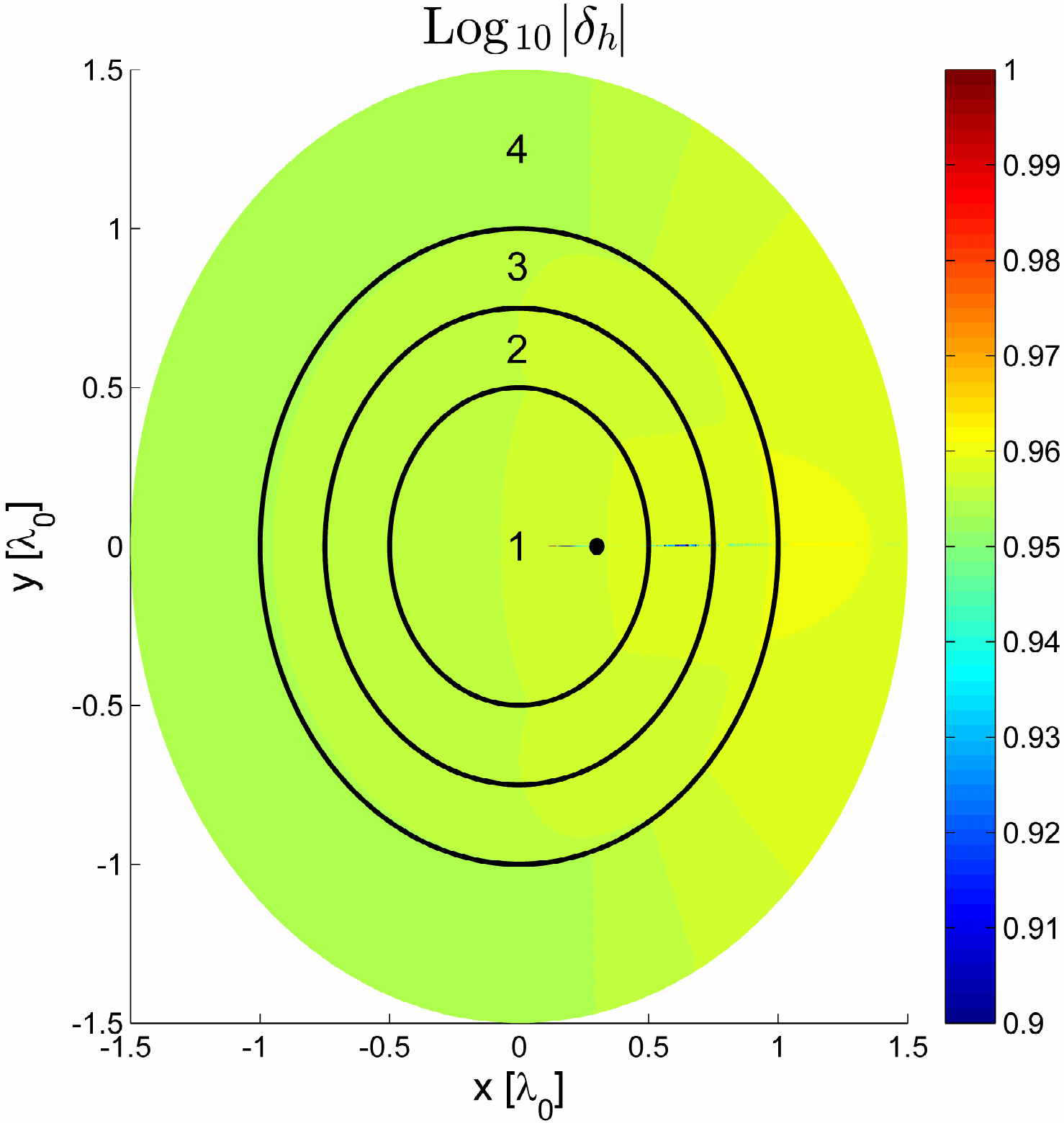}}
\caption{\small Figs. \ref{geom3a} and \ref{geom3b} feature the co-polarized (TM$_z$) field $E_z$ and cross-polarized (TE$_z$) field $H_z$ (resp.) for interface tilting angle $\alpha=1^{\circ}$, and similarly for Figs. \ref{geom3c} and \ref{geom3d} in the $\alpha=10^{\circ}$ case. Fig. \ref{geom3e} shows the relative difference in $E_z$ ($\delta_e$) between the $\alpha=1^{\circ}$ and $\alpha=10^{\circ}$ cases, and likewise for Fig. \ref{geom3f} concerning $H_z$ ($\delta_h$).}
\label{geom3}
\end{figure}

We employ a commercial finite-element solver (COMSOL RF Module) to verify the consequences of inserting annular slabs of ``interface-flattening" T.O. media within geophysical media parameters. Consider a two-layer medium centered at the origin $x=y=0$, with the interface tilted within the $xz$ plane by angle $\alpha=\alpha_1$ (c.f. Fig. \ref{geom2}). Prior to inserting the T.O. media, one has the inner layer (e.g. exploratory borehole) occupying $0\leq \rho < 0.75$m while the outer layer (e.g., background Earth formation) occupies $0.75\mathrm{m} \leq \rho < \infty$. The borehole has material properties $\{\epsilon_1=\epsilon_0,\mu_1=\mu_0,\sigma_1=1\mathrm{S/m}\}$ while the Earth formation has material properties $\{\epsilon_2=\epsilon_0,\mu_2=\mu_0,\sigma_2=2\mathrm{S/m}\}$. After inserting the $d=0.25$m thick annular slabs of T.O. media,\footnote{That is, one slab immediately exterior to, and one slab immediately interior to, the interface at $\rho=0.75$m. The $\rho=0.75$m interface is shown as the solid back circle separating the two layers, marked as ``2" and ``3", in Fig. \ref{geom3}.} one has four concentric, vertically-oriented layers; the resultant outer formation, the annular T.O. slab exterior to the interface at $\rho=0.75$m, the annular T.O. slab interior to the interface at $\rho=0.75$m, and the borehole are labeled as ``4", ``3", ``2", and ``1" in Fig. \ref{geom3} (solid black lines demarcate the layer boundaries). Furthermore, we place a 2 MHz (a typical frequency for logging-while-drilling sondes \cite{anderson1}) electric line source at $(x,y)=(0.3,0)$m (shown as a black dot), which carries a $z$-directed current of 1A

Observing first the cross-polarized fields (Figs. \ref{geom3b}, \ref{geom3d}, and \ref{geom3f}) we notice, for small $\alpha$, that a linear variation in $\alpha$ results in an approximately linear variation in the cross-polarized field levels. This observation is qualitatively confirmed upon comparing Figs. \ref{geom3b} and \ref{geom3d}, and is quantitatively confirmed by Fig. \ref{geom3f}. Indeed, an order of magnitude increase in $\alpha$ (i.e., from $\alpha=1^{\circ}$ to $\alpha=10^{\circ}$) leads to an order of magnitude relative difference in $H_z$ ($\delta_h$) between the two tilt cases. This is shown on log scale by the global, nearly uniform value of Log$_{10}|\delta_h|\sim 1$ in Fig. \ref{geom3f}. This variation in cross-polarized fields versus $\alpha$ is expected if one views the T.O. medium properties as a perturbation on the respective ``host" media; in fact, the T.O. medium properties become that of the respective host medium in the limit as $\alpha \to 0^{\circ}$ (c.f. \eqref{lambda}). Furthermore, observe that $H_z$ is most intense in the neighborhood near $\phi=\pm 90^{\circ}$ and least intense near $\phi=\{0^{\circ},180^{\circ}\}$ (in fact, one observes a null along $y=0$). This trend agrees with our earlier prediction about an azimuthal-dependent impedance mismatch along interfaces bordering T.O. media and their hosts.

Observing the co-polarized field distribution, Fig. \ref{geom3e} suggests on the order of 0.01\% (or $-4$ on the log scale) to 1\% (or $-2$ on the log scale) change in $E_z$. Furthermore we notice that near the source, and particularly in the ``forward" region emanating from the source into the $\phi=0^{\circ}$ direction,\footnote{Note that this would be the zone less susceptible to any mismatch effects.} not only are the cross-polarized fields minimized but there is relative variation in $E_z$ on the order of 0.1\%. Based on these two observations we conclude that this local variation in $E_z$ versus $\alpha$, within the ``forward'' region, is primarily the result of interface tilting and not artificial T.O.-based modeling ``noise". These observed features are also of interest for suggesting a potential route to {\it retrieve} the (local) direction, and magnitude, of tilt observed on a given section of a logging tool by monitoring such (local) azimuthal field variations.

\section{\label{conc}Conclusions}
We proposed a new computational modeling strategy to model EM fields radiated in tilted cylindrically-stratified media. The proposed approach incorporates cylindrical interface tilting through the use of T.O.-prescribed artificial layers. The spatial homogeneity of the T.O.-derived media, which facilitates their inclusion into standard pseudo-analytical algorithms that handle multilayered cylindrical media, as well as their explicit material tensor representations, were mathematically exhibited, and the media were incorporated into 2-D numerical simulations to characterize their behavior. The results favorably suggests its further study as a tool to characterize interface tilting in geophysical borehole sensing.

\section*{\label{conc}Acknowledgment}
The authors acknowledge Mr. Justin Burr and Mr. Zeeshan Zeeshan of the OSU/ESL for assistance in producing the numerical results.
\bibliography{reflist}
\bibliographystyle{IEEEtran}
\end{document}